\documentclass[smallcondensed]{svjour3}     
\smartqed  
\usepackage{amsmath,amssymb,amsfonts}
\usepackage{xAlgorithm}
\usepackage{cite}
\usepackage{wrapfig}
\usepackage{amsmath}
\usepackage{color}
\usepackage{graphicx}
\usepackage{subfigure}
\usepackage{textcomp}
\usepackage{xcolor}
\usepackage{booktabs} 

\begin{document}
\title{Multiset Synchronization with Counting Cuckoo Filters
}

\author{Shangsen Li         \and
        Lailong Luo            \and
        Deke Guo             
}

\institute{Science and Technology on Information System Engineering Laboratory,\\
National University of Defence Technology, Changsha Hunan 410073, P.R. China.\\
\email{lishangsen/luolailong09/guodeke@nudt.edu.cn}  
}

\date{Received: date / Accepted: date}
\maketitle

\begin{abstract}
Set synchronization is a fundamental task in distributed applications and implementations. Existing methods that synchronize simple sets are mainly based on compact data structures such as Bloom filter and its variants. However, these methods are infeasible to synchronize a pair of multisets which allow an element to appear for multiple times. To this end, in this paper, we propose to leverage the counting cuckoo filter (CCF), a novel variant of cuckoo filter, to represent and thereafter synchronize a pair of multisets. The cuckoo filter (CF) is a minimized hash table that uses cuckoo hashing to resolve collisions. CF has an array of buckets, each of which has multiple slots to store element fingerprints. Based on CF, CCF extends each slot as two fields, the fingerprint field and the counter field. The fingerprint field records the fingerprint of element which is stored by this slot; while the counter field counts the multiplicity of the stored element. With such a design, CCF is competent to represent any multiset. After generating and exchanging the respective CCFs which represent the local multisets, we propose the query-based and the decoding-based methods to identify the different elements between the given multisets. The comprehensive evaluation results indicate that CCF outperforms the counting Bloom filter (CBF) when they are used to synchronize multisets, in terms of both synchronization accuracy and the space-efficiency, at the cost of a little higher time-consumption.
\end{abstract}


\section{Introduction}
Consider a couple of hosts $host_A$ and $host_B$ with set $A$ and $B$ respectively, set synchronization means to derive out the elements in $ A\cup B\mathrm{-}A\cap B$ and then exchange them, such that eventually $A\mathrm{=}B\mathrm{=}A\cup B$. Set synchronization is a fundamental task in distributed applications and implementations. For example, in Gossip protocols\cite{minsky2003set}, random pairs of nodes synchronize their content to realize the eventual consistency. In Peer-to-Peer (P2P) networks\cite{eppstein2011s}, two hosts need to get a union of their sets which could be file blocks or link state packets. In personal cloud storage systems such as Dropbox\cite{drago2012inside}, the data synchronization among personal digital devices is implemented as multiple two-party set synchronization tasks. In software-defined networks (SDN)\cite{schiff2016band}, in order to guarantee consistency of network operations, forwarding policies performed on the data plane by different controllers need to be synchronized. In distributed file systems\cite{roy2019managing}, data consistency must be guaranteed during asynchronous updates between the heterogeneous or homogeneous storage systems.

Nowadays, more and more applications rely on the multiset abstraction rather than the previous simple set abstraction. In a multiset, an element can have multiple replicas; by contrast, a simple set element only allows one element instance. For example, in network monitoring, the flows are modelled as multiset elements whose contents are the source and destination IP addresses and multiplicities are their number of packets. In online shopping, to evaluate the popularity of the commodities, customer behaviors are formed as multiset whose element content is the commodity ID and the multiplicity is the user' visiting frequency of the article. In biology, biological evolution and chemical reactions are abstracted as the evolution of multiset and elements' interactions in a multiset object space. This methodology\cite{krishnamurthy2004rule} guarantees the plausibility of DNA computing and programmable living machines. In blockchain, efficient set synchronization protocols are applied to synchronize newly authored transactions and newly mined blocks of validated transactions\cite{ozisik2017graphene}. Such protocols upgrade critical performance and reduce bandwidth consumption significantly in the blockchain context.

Existing methods that synchronize simple sets are mainly based on the compact data structures. The insight is to employ Bloom filter (BF)\cite{bloom1970space} and its variants to provide a content summarization of the local set. After exchanging and comparing these data structures, the different elements can be determined accordingly. Specifically, the counting Bloom filter (CBF)\cite{fan2000summary}, compressed Bloom filter\cite{mitzenmacher2002compressed}, invertible Bloom filter (IBF)\cite{eppstein2011s} and invertible Bloom lookup table (IBLT) \cite{goodrich2011invertible} represent all elements in a set with a vector of cells, each of which can be one single bit or a field with multiple bits. After exchanging these data structures, the different elements are determined by either a query-based (BF and CBF) or a decoding-based (IBLT and IBF) mechanism. Consequently, it is not necessary to transfer the common elements.

However, the above simple set synchronization methods are not feasible to multisets. First, they may fail to represent the multiplicity information of multiset members. The original BF bit vector definitely cannot record elements whose multiplicities are larger than 1. The XOR operations in IBF and IBLT, however, disable the representation of multiset elements since the elements with even multiplicities will be eliminated from the multiset. Second, these methods may fail to distinguish the different elements caused by distinct content ($d_E$) from the different elements due to unequal multiplicities ($d_M$). For example, CBF represents multiset elements with its counters in its cell vector naturally. However, the query-based mechanism cannot distinguish $d_E$ from $d_M$. Generally, only the elements in $d_E$ are required to be transferred, while the elements in $d_M$ are synchronized via generating dedicated number of replicas. Consequently, distinguishing them is quite important for bandwidth-scarce scenarios.

In this paper, we present a novel design of cuckoo filter\cite{fan2014cuckoo}, namely the counting cuckoo filter (CCF), and thereafter propose multiset synchronization methods based on this data structure. Based on the original cuckoo filter, we attach a new counter field in each slot to record the multiplicity of the stored element. Specifically, a CCF consists of a cuckoo hashing table with $b$ buckets. Each bucket contains $w$ slots and can accommodate at most $w$ fingerprints. In each slot, there are two fields, i.e., the fingerprint and counter, which record the fingerprint of element mapped into that bucket and its multiplicity, respectively. CCF represents multiset members by storing their fingerprints and multiplicities directly. The insertion of elements also follows the \emph{kick-and-reallocate} strategy introduced in \cite{fan2014cuckoo}.

After receiving the CCF from the other host, the local host compares the received CCF with its own, and determines the elements in $d_E$ and $d_M$, respectively. By traversing the elements in the root set, local host query the multiplicity information of elements in the received CCF. By jointly considering the existence and multiplicity information, the local host can effectively distinguish $d_E$ from $d_M$. The host can also directly compare the two CCFs and eliminate common elements. By decoding the labelled slots in the decoded CCF, local host can correctly identify the corresponding elements' content and classify the different elements into $d_E$ and $d_M$ respectively. The elements in $d_M$ can be replicated locally for synchronization. Consequently, only the elements in $d_E$ will be transmitted to the other host.

The major contributions of this paper can be summarized as follows:
\begin{itemize}
\item We formulate the design of CCF and analyze the false positive rate of it. In conclusion, CCF is more space efficient than CBF in theory when representing multisets.
\item We design two multiset synchronization methods based on CCF. By querying the CCF received from the other host, the local host determines elements in $d_E$ and $d_M$ reasonably. The local host can also eliminate the common elements and decode the different elements in $d_E$ and $d_M$ correctly and efficiently.
\item Comprehensive experiments conclude that the CCF-based method outperforms the CBF-based method in terms of synchronization accuracy and space overhead. Besides, compared with the decoding-based method, the query-based method is more effective in addressing multiset synchronization problem in terms of accuracy and time-consumption.
\end{itemize}
\par The remainder of this paper is organized as follows. Section 2 presents related preliminaries. Section 3 introduces the design of CCF and its associated operations to represent multisets. Section 4 details the methodology of multiset synchronization enabled by CCF. Section 5 theoretically compares CCF and CBF in terms of false positive rate and space efficiency. Section 6 reports the evaluation results about our CCF-based method in terms of time-consumption, synchronization accuracy and space overhead. Finally, Section 7 concludes this paper.

\section{Preliminaries}
In this section, we introduce the related compact data structures, i.e., Cuckoo filter and counting Bloom filter.

\subsection{Cuckoo filter}
A Cuckoo filter (CF) is a hash table with $b$ buckets, each bucket has $w$ slots to accommodate at most $w$ elements. When inserting a key, cuckoo hashing table applies two perfectly random hash functions to determine the two possible bucket entries where the key can be inserted\cite{richa2001power}. Unlike the traditional cuckoo hashing table, which stores the element content directly, Fan \emph{et al}.\cite{fan2014cuckoo} represent the elements by recording their fingerprints instead in cuckoo filter. Cuckoo filter further applies the partial-key strategy to determine the candidate buckets during the insertion phase and locate the alternative candidate bucket during the reallocation phase. Specifically, for an element $x$, the partial-key hashing mechanism calculates the index of the two candidate buckets with the following two functions:
\begin{equation}
\begin{aligned}
&h_1\mathrm{=}hash\left( x \right)\\
&h_2\mathrm{=}h_1\oplus hash\left(\eta_x\right)\\
\label{bucket}
\end{aligned}\end{equation}
\vspace{-3ex}

To represent an element $x$, if either of its candidate bucket is available, the element fingerprint will be stored in a random slot there. If there is no any empty slot for $x$, then CF randomly selects a stored fingerprint as victim. The victim is then kicked out from its accommodation to store the fingerprint of $x$ instead. Then the victim will be reallocated to its alternative candidate bucket. During the reallocation, based on Equ. \ref{bucket}, the index of the alternative candidate bucket is determined by XORing its current position index with the hash value of its fingerprint. This partial-key strategy speeds up the reallocation process significantly. The reallocation terminates when there is no further victim or the number of reallocations reaches a given threshold $max$. When an element is failed to be inserted, the CF is declared as \emph{'FULL'}.

CF is elegant to represent simple set, but is inefficient to represent multiset. The reason is that representing each element replicas separately with a CF slot aggressively occupies the space. Such a space-inefficient method is surely not advisable. Moreover, when the value of $m_X(x)$ is larger than $2w$, the CF cannot tell the correct multiplicity information of $x$.

\subsection{Counting Bloom filter}
CBF \cite{fan2000summary} is a known variant of Bloom filter\cite{bloom1970space} which leverages the $k$ bits in a vector to represent the absence and existence of an element in a set. Despite of the constant-time complexity and the space-efficiency features, Bloom filter cannot support the deletion of elements. The reason is that resetting the $k$ corresponding bits from 1s to 0s may cause the mis-deletion of other elements which are also mapped to these bits. To this end, CBF replaces each bit in the vector with a cell (counter) with multiple bits. Whenever an element is mapped into cells, the $k$ counters will be added up by 1. Then the deletion of elements is straightforwardly realized by decreasing the $k$ corresponding counters by 1. Consequently, the deletion of element $x$ will not affect the membership information of other elements.

As stated in \cite{luo2017efficient}, it is possible to realize multiset synchronization with CBF. The core idea is to represent each multiset as a CBF vector, same as the Count-min sketch\cite{cormode2005improved}, the minimum value among the $k$ counters in CBF is regarded as the multiplicity of an elements. By subtracting the two generated CBFs, the different elements can be derived out by querying the local set against the subtracted result. However, this method incurs unacceptable synchronization accuracy. To achieve higher synchronization accuracy, the CBFs should be lengthened. The space overhead can be tremendous to achieve expected synchronization accuracy. Besides, one counter may be increased by the insertion of multiple elements, so that it can overflow easily. To avoid this overflow, CBF needs to augment more bits to the counter field, which further leads to more space overhead\cite{rottenstreich2014variable}. Therefore, counting bloom filter is incompetent to represent and synchronize multiset. Consequently, this paper presents the counting cuckoo filter data structure, which represents and thereby synchronizes multisets elegantly.

\section{The CCF design for multiset representation}

\subsection{Multiset: prior knowledge and current representation}
As the general concept of the simple set, multisets allow elements to appear for more than once \cite{luo2017efficient}. For any element $x$ in a multiset $X$, it associates with an integer multiplicity $m_X\left(x\right) $ to explicitly indicate the number of its replicas in multiset $X$. $X$ has a root set $X^*$ which contains the elements appear in multiset $X$ regardless of their multiplicities. Let $C(X) \mathrm{=} \sum_{x\in X} m_X(x)$ denotes the cardinality of multiset $X$, which is the sum of multiplicities of its elements \cite{blizard1991development}. With the above notations, a multiset can be represented by the elements in root set and the multiplicity information of them.

\begin{figure}[htbp]
\centerline{\includegraphics[width=3in]{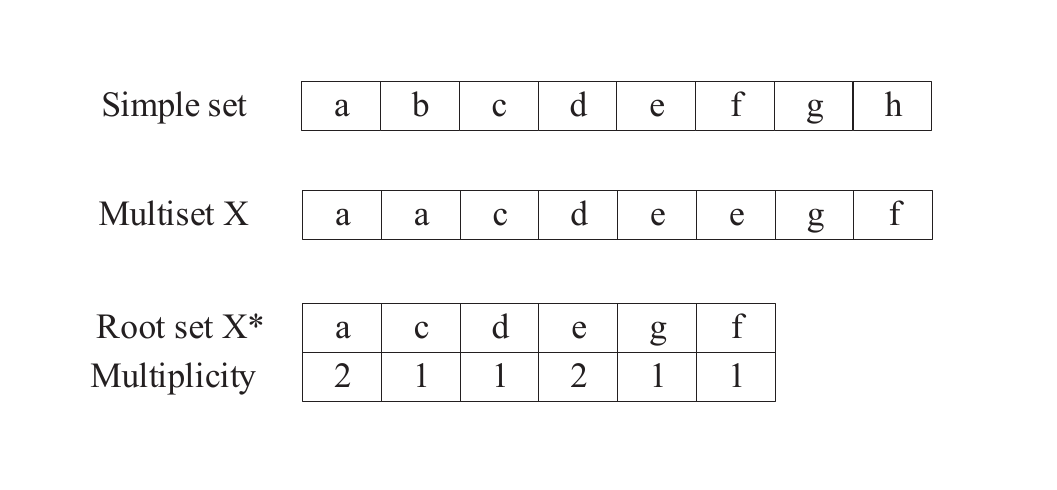}}
\caption{Simple set and Multiset}
\label{multiset}
\end{figure}

As illustrated in Fig. \ref{multiset}, unlike simple set, multiset elements can appear for more than once. The total number of elements in the multiset is 8, so the cardinality of the multiset $C(X)$ is 8. For the element $a$, its multiplicity $m_X\left(a\right) $ is 2. The multiset $X$ can be represented by its root set $X^*$ which records the elements appear in the multiset and the corresponding multiplicity information jointly.

The existing methods used to represent multiset are mainly based on Bloom filters. Specifically, the BF-based methods represent the root set of multiset iteratively. This is an effective way to represent and synchronize multiset. But in the worst case, the rounds can be the maximum multiplicity of elements in multiset\cite{luo2017efficient}. As a consequence, this method may cause unacceptable computation overhead. The multiset can also be represented by the counting-based method. It is feasible to represent the multiplicity information with the counting-based method. CBF leverages a number of counters to represent a single element. To get the targeted false positive rate, CBF needs to increase the number of counters. This strategy incurs much inefficient space overhead.

Intuitively, CF is also a compact data structure which can represent simple sets. CF can support a small number of replicas of some items. If each bucket can accommodate 4 fingerprints, the cuckoo filter can support up to 8 replicas at most. Although it isn't clear how the insertion of replicas will impact the probability of failure during insertion of other items\cite{pandey2017general}. However, undoubtedly, storing the multiset elements' replicas directly with the CF candidate buckets damages the space efficiency. Besides, by doing so, the insertion failure comes early, which leads to a lower space occupancy ratio. The newly variant of CF, i.e., the dynamic cuckoo filter (DCF)\cite{chen2017dynamic} is possible to represent multiset elements by adaptively resizing its capacity. But it may incur much space overhead because of the skewed data distribution. Specifically, the multiplicities of some elements are much larger than others, DCF must augment more sparse CFs to accommodate hot items in multisets. These added CF vectors are inefficient.

We argue that the counting-based data structure is efficient to represent multiset. And the $counter$ field is advisable to record the multiplicity information. To this end, we design our CCF data structure and the related operations to address the representation and synchronization problem. To ensure the targeted false positive rate, CCF only need to increase the bits of fingerprint. Compared with CBF, CCF incurs much less space overhead.

\subsection{The CCF data structure design}
We found a real implementation of CCF by Barrus \footnote{https://pyprobables.readthedocs.io}. However, formalized description and analysis are still missing.  Therefore, in this subsection, we present a detailed description of the CCF data structure, as well as its associated functionalities to represent multisets.
\begin{figure}[htbp]
\centerline{\includegraphics[width=3.40in]{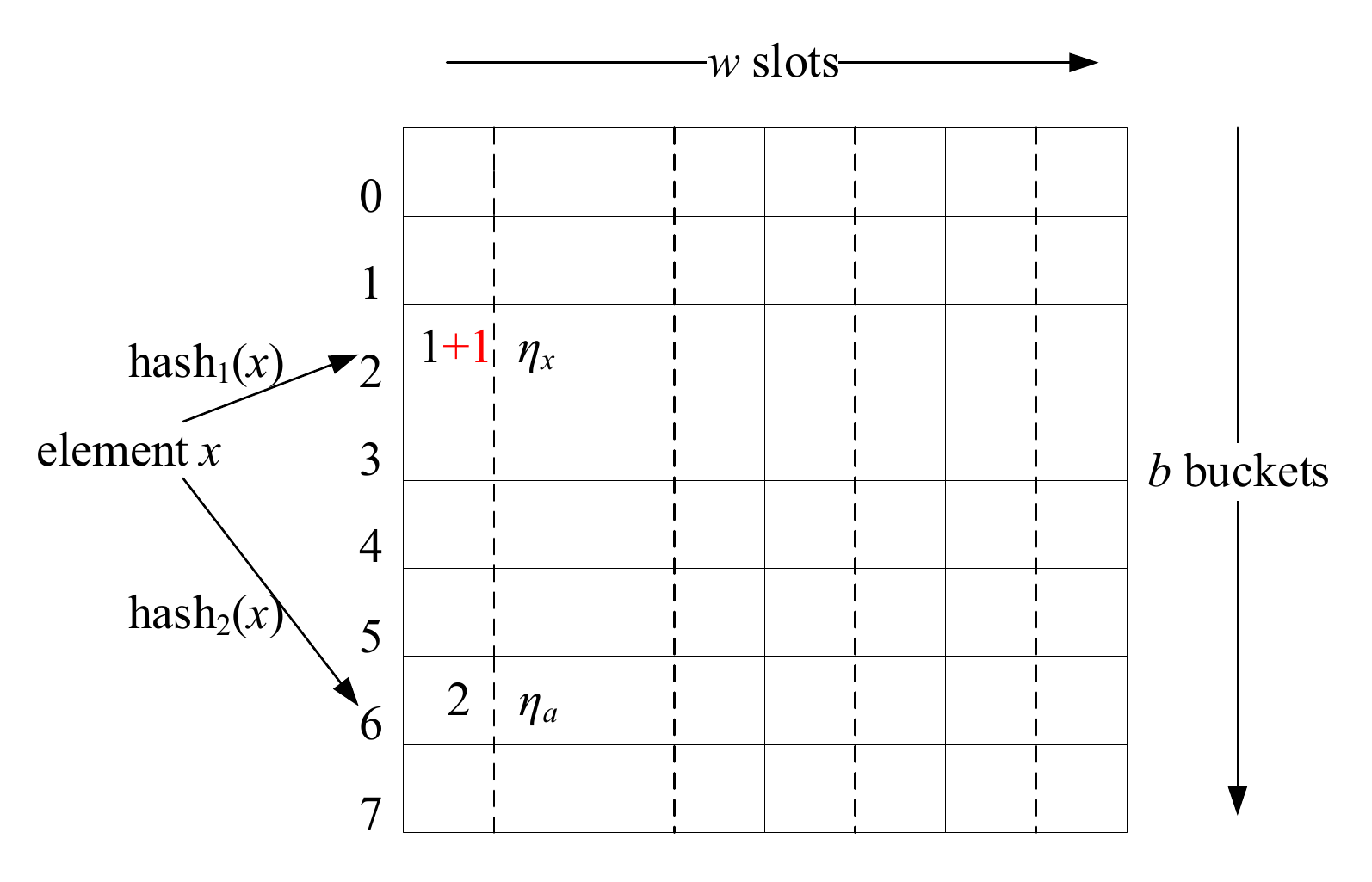}}
\caption{Structure of counting Cuckoo filter}
\label{structure}
\vspace{-1ex}
\end{figure}

As plotted in Fig. \ref{structure}, a CCF consists of $b$ buckets, each of which has $w$ slots to accommodate at most $w$ element fingerprints. In addition, each slot contains two fields, i.e., the fingerprint field $\eta$ and the $counter$ field. The fingerprint field is responsible to record the fingerprint of element which are stored into this slot; while the $counter$ field identifies the multiplicity of that element. With such a design, CCF can naturally represent multiset members with the following insertion, deletion and query operations.

\textbf{Insertion.} CCF still records element fingerprints by following the \emph{kick-and-reallocate} strategy to represent multisets. Initially, for an arbitrary multiset element $x$, if either of its candidate bucket has an empty slot, the fingerprint $\eta_x$ and its multiplicity will be stored there directly (Line 2 to 5). Otherwise, a random fingerprint and its multiplicity in these two candidate buckets are kicked out to accommodate element $x$. The victim is thereafter reallocated to its alternative candidate bucket with the partial-key strategy. The index of the alternative bucket can be directly obtained by XORing current bucket index with the hash of the fingerprint. This reallocation ends successfully when there is no further victim triggered or returns unsuccessfully when the number of reallocations reaches the given threshold $max$ (Line 7 to 19). The value of $max$ is decided with the joint consideration of CCF length, number of elements to represent, acceptable time-consumption and the expected occupancy ratio. The details of insertion are given in following Algorithm 1.

\begin{algorithm}[]
\small
\caption{Insertion($x$) at $host_A$}\label{alg:Insert}
\SetKwData{Temp}{cand}
\SetKwData{Stat}{state}
\Input{Element $x$, $CCF_A$}
\Output{Insertion state}
\State{Calculate $h_1,h_2,\eta_x$ of element $x$}
\If{existing empty $slot$ in $bucket_{h_1}$ or $bucket_{h_2}$\\}{
\State{store $\eta_x$ and $m_A(x)$ with that empty slot}
\RETURN{True}
}
\Else{
\State{Randomly select $slot$ in $bucket_j(j\in{h_1,h_2})$}
\State{ $\eta_k$=$slot.\eta$, $count_k$=$slot.counter$}
\State{store $\eta_x$ and $m_A(x)$ in this slot}
\While{$i\mathrm{<}max$} {

\State{$bucket_{alt}\mathrm{=}j\oplus hash(\eta_k)$}
\If{bucket $bucket_{alt}$ has an empty slot}{
\State{store $\eta_k$ and $count_k$ in this slot}
\State {\RETURN{True}}
}
\Else{$i\mathrm{++}$\\
\State{Randomly select $slot$ in $bucket_{alt}$}
\State{ $\eta_k$=$slot.\eta$, $count_k$=$slot.counter$}}}
\RETURN{False}}
\end{algorithm}

\textbf{Deletion}. As specified in Algorithm 2, to delete an element $x$, CCF first calculates its fingerprint $\eta_x$ and its two candidate buckets. If $\eta_x$ can be found in neither of these candidate buckets, the algorithm returns \emph{False} to demonstrate that $x$ is not represented by CCF (Line 6). Otherwise, in the target slot, the fingerprint field is cleared to empty and the counter field is reset to 0 (Line 4 to 5). This deletion strategy can eliminate multiple replicas directly. The time-complexity of deletion is constant, since only two buckets are checked.

\begin{algorithm}[]
\small
\caption{Delete($x$,$CCF$)}\label{alg:Delete}
\SetKwData{Temp}{cand}
\SetKwData{Stat}{state}
\Input{Element $x$,$CCF$}
\Output{Deletion state}
\State{Calculate $h_1,h_2,\eta_x$ of element $x$}
\If{existing $slot.\eta\mathrm{=}\eta_x$ in $bucket_{h_1}$ or $bucket_{h_2}$}{
\State{$slot.\eta\mathrm{=}Null$, $slot.counter\mathrm{=}0$}
\RETURN{True}}
\Else{\RETURN {False}}
\end{algorithm}

\begin{algorithm}[]
\small
\caption{Query($x$,$CCF$)}
\SetKwData{Temp}{cand}
\SetKwData{Stat}{state}
\Input{Element $x$$, CCF$}
\Output{Multiplicity of element $x$}
\State{Calculate $h_1,h_2,\eta_x$ of element $x$}
\If{existing $slot.\eta\mathrm{=}\eta_x$ in $bucket_{h_1}$ or $bucket_{h_2}$}{
\State{$multiplicity\mathrm{=}slot.counter$}}
\Else{$multiplicity\mathrm{=}0$}
\RETURN$multiplicity$
\end{algorithm}

\textbf{Query.} Algorithm 3 specifies the pseudocode of querying an arbitrary element $x$. Basically, querying a multiset element means to determine its membership, as well as its multiplicity. Straightforwardly, CCF checks the two candidate buckets of a given element $x$. If the fingerprint can be found in either of them, CCF returns the count field of that slot to indicate that $x$ is a member of the represented multiset, and its multiplicity is exactly the returned value (Line 4 to 5). Otherwise, CCF returns 0 to declare that $x$ is not a member of the represented multiset (Line 6). The time-complexity of query is also constant.

\section{Multiset synchronization with CCF}

In this section, we propose two multisets synchronization methods based on the above CCF data structure. Before specifying the synchronization methods, we first introduce the framework for addressing such multiset synchronization problems.

\subsection{The synchronization framework}
\begin{figure}[t]
\centerline{\ \ \ \ \includegraphics[width=3.40in]{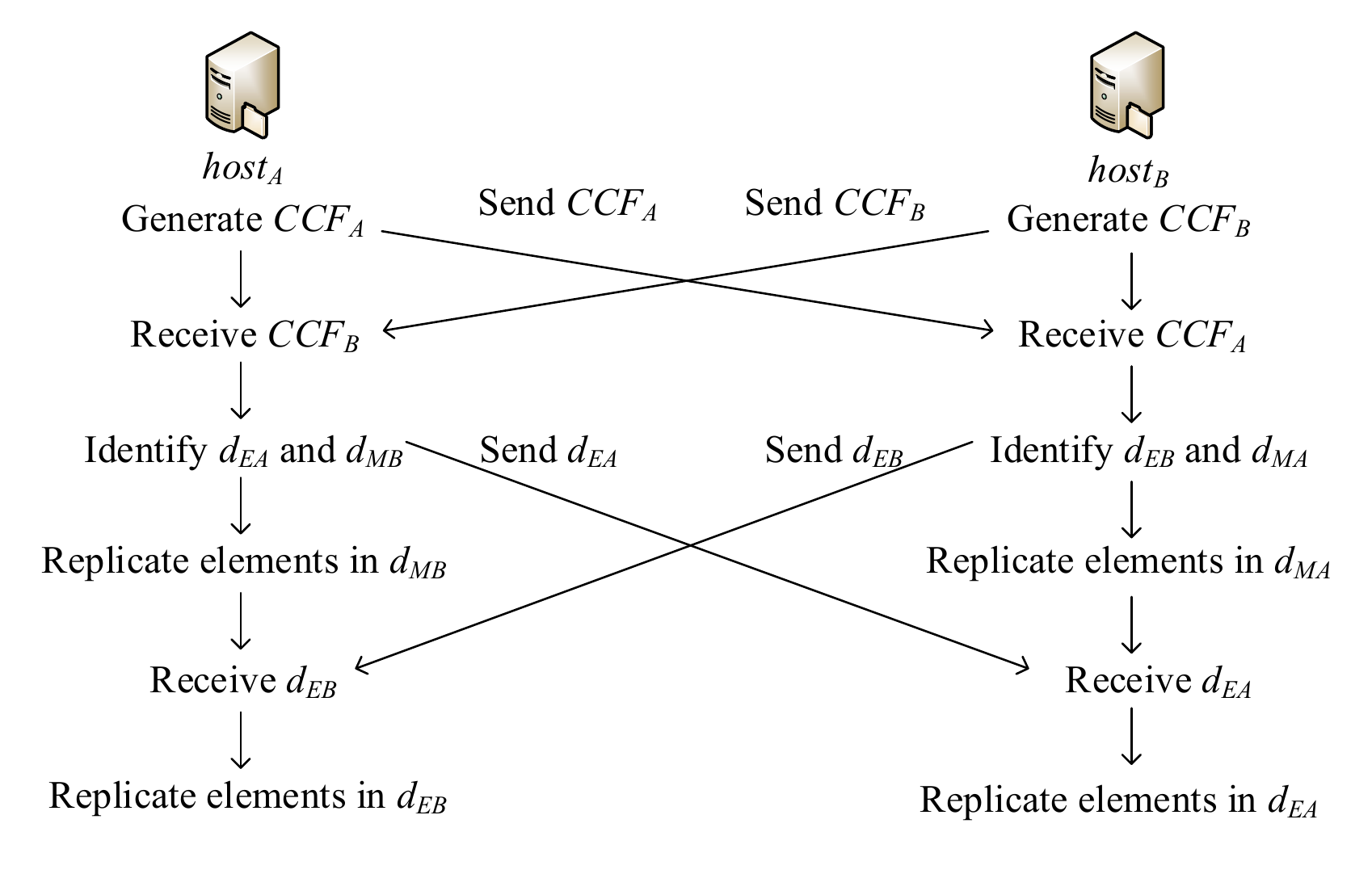}}
\caption{The framework of multiset synchronization}
\label{framework}
\end{figure}

As shown in Fig. \ref{framework}, to synchronize a pair of multisets, there are two rounds of interactions in the synchronization process between two hosts. In the first round of interaction, the hosts exchange their local CCFs; while the second round of interaction is to transmit the discovered different elements in $d_E$. The details of the synchronization framework are specified as follows.

First of all, $host_A$ and $host_B$ represent their local multisets with the CCF data structure as $CCF_A$ and $CCF_B$, respectively. After that, $host_A$ sends its $CCF_A$ to $host_B$ and vice versa. After such an interaction, both $host_A$ and $host_B$ acquire the information of the other multiset.

Then the task is to deduce the different elements at each host. Let $d_{E_A}$ and $d_{E_B}$ denote the elements that only contained in the multiset $A$ and $B$, respectively. Pair-wisely, let $d_{M_B}$ and $d_{M_A}$ denote the elements such that $m_B(x)\mathrm{>} m_A(x)$ and $m_A(x)\mathrm{>} m_B(x)$, respectively. Then $host_A$ should identify the elements in both $d_{E_A}$ and $d_{M_B}$, such that the elements in $d_{E_A}$ will be thereafter sent to $host_B$ and the elements in $d_{M_B}$ will be replicated locally for the synchronization purpose. Note that, the number of further generated replicas of elements in $d_{M_B}$ is $m_B(x)\mathrm{-} m_A(x)$ at $host_A$. Similarly, $host_B$ needs to identify the elements in $d_{E_B}$ and $d_{M_A}$. Then the elements in $d_{E_B}$ will be transferred to $host_A$ while  the elements in $d_{M_A}$ will be replicated locally such that $m_A(x)\mathrm{=} m_B(x)$. After receiving $d_{E_B}$ from $host_B$, $host_A$ generates dedicated number of replicas according to the multiplicity information and $host_A$ also does the same thing to complete the synchronization process.

By following the above framework, only the elements in $d_E$ will be transmitted for once. The elements in $d_M$, on the contrary, will not be transmitted. This communication-friendly synchronization framework is quite important for bandwidth-scarce scenarios. According to the above framework, the core problem of multiset synchronization with CCF is how to deduce the different elements, in both $d_E$ and $d_M$. In this paper, we state that our CCF data structure enables both the query-based method and the decoding-based method to identify different elements between multisets $A$ and $B$. The details are given as follows.

\subsection{Identify the different elements via querying}

After receiving $CCF_B$ from $host_B$, $host_A$ can deduce the different elements by simply querying its local elements in its root set $A^*$ against $CCF_B$. As stated in Section 3, CCF responds the query by returning either 0 to indicate that the queried element is not stored by this CCF or the exact multiplicity of queried element. Therefore, the joint consideration of the query result and the local element information will tell whether the queried element is a different element or not.

With the above insight, Algorithm 4 details our query-based method. Note that, this query-based method is capable of distinguishing elements in $d_E$ from those in $d_M$. If the query result is zero, $x$ is not a member of multiset $B$ and should be added into $d_{E_A}$ for later transmission (Line 3 to 4). On the other hand, if the query result is more than zero, it means that $x$ is also a member of multiset $B$ and the $m_B(x)$ equals to the returned value. Especially, if $m_A(x)$ is equal to $m_B(x)$, it implies that the element $x$ shares the same multiplicity in both multisets, no further action is triggered. On the contrary, if  $m_B(x)$ is larger than $m_A(x)$, element $x$ is identified as a member of $d_{M_B}$ (Line 5 to 7). According to the framework presented above, additional $m_B(x)\mathrm{-}m_A(x)$ replicas of $x$ will be generated at $host_A$ such that eventually $m_B(x)\mathrm{=}m_A(x)$.

\begin{algorithm}[t]
\small
\caption{Identifying different elements with the query-based method at $host_A$}
\SetKwData{Temp}{cand}
\SetKwData{Stat}{state}
\Input{Root set $A^*$ of multiset $A$, $CCF_B$}
\Output{Elements in $d_{E_A}$ and $d_{M_B}$ with multiplicity information}

\For{each element $x$ in root set $A^*$}{
\State{$m_B\left( x \right)\mathrm{=}query\left( x,CCF_B \right)$}
\If{$m_B\left( x \right)\mathrm{=}0$}{
\State{add $x$ and $m_A(x)$ into $d_{E_A}$}}
\Else {
\If{$m_A\left( x \right)\mathrm{<}m_B\left( x \right)$}{
\State{add $x$ and $m_B(x)\mathrm{-}m_A(x)$ into $d_{M_B}$}}}}
\end{algorithm}

Pair-wisely, $host_B$ queries the elements in its local root set $B^*$ against the received $CCF_A$. The query result derives elements in $d_{E_B}$ and $d_{M_A}$ respectively.

\begin{algorithm}[t]
\small
\caption{Identifying different elements with the decoding-based method at $host_A$}
\SetKwData{Temp}{cand}
\SetKwData{Stat}{state}
\Input{Root set $A^*$ of multiset $A$, $ CCF_A$, $CCF_B$}
\Output{Elements in $d_{E_A}$ and $d_{M_B}$ with multiplicity information}
\For{$i\mathrm{<}b$}{
\For{$j\mathrm{<}w$}{
\If{$CCF_A [i][j].\eta$ is not empty}{
\If{$CCF_A [i][j].\eta$ in $slot$ of $CCF_B$}{
\State{$count\mathrm{=}CCF_A [i][j].count$}
\If{$count\mathrm{<}slot.count$}{
\State{$CCF_A [i][j].flag\mathrm{=}1$}
\State{$CCF_A [i][j].count\mathrm{=}slot.count\mathrm{-}count$}}
\Else{clear the slot $CCF_A [i][j]$}}
\Else{$CCF_A [i][j].flag = 0$}}}}
\For{each element $x$ in root set $A^*$}{
\If{$m_A\left( x \right)\mathrm{>}0\ $ and $flag\mathrm{=}0$}{
\State{add $x$ and $m_A(x)$ into $d_{E_A}$}}
\Else {
\If{$m_A\left( x \right)\mathrm{>}0\ $ and $flag\mathrm{=}1$}{
\State{add $x$ and $m_A(x)$ into $d_{M_B}$}}}}
\end{algorithm}
\vspace{-1ex}

\subsection{Identify the different elements via decoding}
In this subsection, we specify how the decoding-based method eliminate the common elements and determine the different elements in $d_{E}$ and $d_{M}$.

After receiving $CCF_B$ from $host_B$, $host_A$ decodes the different elements in $d_{E_A}$ and $d_{M_B}$ with two main steps. The first step is to eliminate the common elements from $CCF_A$ and label slots which hold the different elements. The second step is to determine the exact element content of the labelled slots in $CCF_A$. To this end, one more bit, i.e., the flag bit, is added into each slot in $CCF$. As specified in Algorithm 5, for each non-empty slot in $CCF_A$, we try to search out the stored fingerprint in the two corresponding buckets in $CCF_B$ (Line 4 to 12). If we fortunately find this fingerprint in the $slot$ of $CCF_B$ and $slot.count$ is larger than $CCF_A[i][j].count$, the flag bit in $CCF_A[i][j]$ is set to 1 to explicitly indicate that this fingerprint corresponds to an element in $d_{M_B}$ (Line 6 to 8). On the other hand, if $slot.count$ is not larger than $CCF_A[i][j].count$, we empty the $CCF_A[i][j]$ slot (Line 10). Unfortunately, if the fingerprint $CCF_A[i][j].\eta$ can be found in neither of the two candidate buckets of $CCF_B$, we set $CCF_A[i][j].flag$ value as 0 to signify that this element belongs to $d_{E_A}$ (Line 12).

After the above elimination and labelling process, $CCF_A$ only stores the fingerprints corresponding to elements in $d_{E_A}$ and $d_{M_B}$. Then Algorithm 5 can easily decode them out by traversing the CCF vector and the local multiset (Line 13 to 18). Similarly, $host_B$ also performs the same algorithm to decide the elements in $d_{E_B}$ and $d_{M_A}$. Surely, following the synchronization framework, $host_A$ will send $d_{E_A}$ to $host_B$ and $host_B$ will transmit $d_{E_B}$ to $host_A$. The elements in $d_{M_A}$ and $d_{M_B}$ are synchronized by generating dedicated number of replicas locally. Such that eventually $m_B(x)\mathrm{=}m_A(x)$ for any element in $host_A$ and $host_B$.

\section{Performance analysis}
In this section, we conduct a theoretical analysis of our data structure CCF and the proposed synchronization methods, by using CBF as a reference.
\subsection{Space efficiency and false positive rate}
BF leverages ${\frac{m}{n}\ln 2}$ bits\cite{luo2018optimizing} for each element to achieve high space efficiency. To realize the target false positive rate $\epsilon$, the parameters of BF, ${i.e.}$, the length of BF bits array $m$, the number of employed hash functions $k$, and the number of elements in simple set $n$, need to be carefully designed. Theoretically, the value of false positive rate can be calculated as follows:
\begin{equation}\begin{aligned}
\epsilon\mathrm{=}\left[ 1\mathrm{-}\left( 1\mathrm{-}\frac{1}{m} \right) ^{nk} \right] ^{k}\approx \left( 1\mathrm{-}e^{\mathrm{-}\frac{kn}{m}} \right) ^k
\label{cbf_false}
\end{aligned}\end{equation}
To reach the minimum $\epsilon$, we can calculate the $k_{op}$ as:
\begin{equation}\begin{aligned}
k_{op}\mathrm{=}\frac{m}{n}\ln 2\approx \frac{9m}{13n}
\label{k}
\end{aligned}\end{equation}

According to $k_{op}$ calculated by Equ. \ref{k}, the resultant false positive rate $\epsilon \mathrm{\approx} 0.5^k\mathrm{\approx} 0.6185^{m/n}$. In other words, to achieve the target false positive rate, the value of $m$ should be increased linearly with the value of $n$. The analysis of false positive rate is also functional in the CBF context. The only difference is that the space overhead of CBF is much more than the original BF. Therefore, to get the false positive rate decreased by $38.15\%$, extra $n$ (the number of elements in the root set) counters are introduced, which means one more counter for each element should be added into the CBF vector. Therefore, for devices with scarce memory, they have to sacrifice the false positive rate of CBF to some extent.

In contrast, we can decrease the false positive rate of CCF by $50\%$ through augmenting one more bit to the fingerprint field of each slot. The details are discussed as follows.

In effect, CCF randomly generates a tuples $(h_{1_x}, h_{2_x}, \eta_x)$ for any inserted element $x$, including the index of two candidate buckets and the fingerprint $\eta_x$. We observe that there are two kinds of hash collisions in CCF. The first kind of collision is caused by complete collision of the two tuples, i.e., the three items contained in the tuples for two distinct elements are equal. For example, for element $x,y$ in multiset $X$, the tuples of elements are $(h_{1_x}, h_{2_x}, \eta_x)$ and $(h_{1_y}, h_{2_y}, \eta_y)$, respectively. If $h_{1_x}\mathrm{=}h_{1_y}, \eta_x\mathrm{=}\eta_y$, and $h_{2_x}\mathrm{=}h_{2_y}$, this collision cannot be identified unless we lengthen the fingerprint to hopefully get different fingerprints for $x$ and $y$. Another kind of collision is caused by the order of candidate bucket indexes. If $h_{1_x}\mathrm{=}h_{2_y}, \eta_x\mathrm{=}\eta_y$ and $h_{2_x}\mathrm{=}h_{1_y}$, the collision can be identified by employing one more bit to mark the order of two candidate buckets. Specifically, if $bucket_{h_1}\mathrm{>}bucket_{h_2}$, set the differentiated bit as 1; if $bucket_{h_1}\mathrm{<}bucket_{h_2}$, remain the differentiated bit as 0; when $bucket_{h_1}\mathrm{=}bucket_{h_2}$, however, this kind of collision cannot be identified when the fingerprint length is not changed.

Then, we discuss the details of the false positive rate caused by the above two kinds of collisions when a multiset is represented by the CCF data structure. And the number of distinct elements of multiset is $N$. A default assumption of the following analysis is that all the multiset elements are inserted successfully into CCFs.

We consider the situation where no differentiated bits are used in CCF. The total number of possible tuples can be estimated as $b\mathrm{\times} 2^f$. Note that, there are $b$ buckets in total and each bucket can choose its random residences from the total $2^f$ fingerprints. For an element $x$, once the corresponding fingerprint $\eta_x$ and one of the candidate bucket $h_{1_x}$ is determined, the other candidate bucket $h_{2_x}$ can be derived accordingly. Thus it can be calculated that the total number of possible combination of tuples is $b\mathrm{\times} 2^f$. Suppose a CF which represents $j\mathrm{-}1$ elements without collisions among these elements. The ratio of tuples which have been mapped by these $j\mathrm{-}1$ elements is  $\frac{j\mathrm{-}1}{b\mathrm{\times}2^f}$. When we insert the $j^{th}$ element into CCF, the ratio of tuples that are available and incur no collision with the existing $j\mathrm{-}1$ elements is estimated as
$1\mathrm{-}2\mathrm{\times}\frac{j\mathrm{-}1}{b\mathrm{\times} 2^{f}}\mathrm{=}1\mathrm{-}\frac{j\mathrm{-}1}{b\mathrm{\times} 2^{f\mathrm{-}1}}$.
Thus the probability that the $j^{th}$ element won't collide with other elements can be calculated as ${1\mathrm{-}\frac{j\mathrm{-}1}{b\times 2^{f\mathrm{-}1}}}$. Thus the false positive rate without the differentiated bit is estimated as follows:

\begin{equation}
\begin{aligned}
&\epsilon\approx1\mathrm{-}\prod_{j\mathrm{=}2}^{N}{\left(1\mathrm{-}\frac{j\mathrm{-}1}{b\mathrm{\times} 2^{f-1}} \right)}\\
\label{false}
\end{aligned}\end{equation}
\vspace{-2ex}

Considering the complexity of Equ. \ref{false}, we propose an approximation to get a lower bound as follows. We note that the probability that the $j^{th} \left(j\geqslant2\right)$  element doesn't collide with the former $j\mathrm{-}1$ elements is less than or equal to the fixed value  $1\mathrm{-}\frac{2}{b\mathrm{\times}2^f}$. Then we reason the lower bound of false positive rate as follows:
\begin{equation}
\begin{aligned}
&\epsilon \geqslant 1\mathrm{-}\left( 1\mathrm{-}\frac{2}{b\mathrm{\times}2^f} \right) ^{N\mathrm{-}1}\approx \frac{2\left( N\mathrm{-}1\right)}{b\mathrm{\times} 2^f}\\
&\ \geqslant \frac{2b\mathrm{\times}w}{b\mathrm{\times}2^f}\mathrm{=}\frac{w}{2^{f\mathrm{-}1}}\\
\label{false_positive}
\end{aligned}\end{equation}
\vspace{-3ex}

In Equ. \ref{false_positive}, we assume that the number of inserted elements equals to the number of slots in the CCF. In other words, we assume that all the elements are inserted successfully into the CCF and all the slots in CCF have elements inserted ideally.

Combine with the conclusions derived from\cite{fan2014cuckoo}, which get the upper bound of the total probability of a false positive hit is:
\begin{equation}\begin{aligned}
\epsilon \leqslant 1\mathrm{-}\left(1\mathrm{-}\text{1/}2^f \right) ^{2w}\approx \frac{w}{2^{f\mathrm{-}1}}
\end{aligned}\end{equation}
we can form the relationship between fingerprint field length $f$ and the target false positive rate $\epsilon$ as follows:
\begin{equation}\begin{aligned}
\epsilon \approx \frac{w}{2^{f-1}}
\end{aligned}\end{equation}

\begin{equation}\begin{aligned}
f\approx \log _2w\mathrm{+}\log _2\frac{1}{\epsilon}\mathrm{+}1
\label{f}
\end{aligned}\end{equation}
We can also estimate the upper bound of collided elements, when the multiset is represented by a CCF, as $N\mathrm{\times }\frac{w}{2^{f\mathrm{-}1}}\mathrm{=}b\mathrm{\times}w\mathrm{\times} \frac{w}{2^{f\mathrm{-}1}}$.

Consider the situation where one more bit is employed in each slot as the differentiated bit. To avoid the first kind of collision, when inserting the $j^{th}$ element of multiset into CCF, the ratio of tuples that remain unused is ${ 1\mathrm{-}\frac{j-1}{b\mathrm{\times}2^f} }$. And the second kind of collision is eliminated by the differentiated bit. In this scenarios, the value of $\epsilon$ is calculated as follows:

\begin{equation}\begin{aligned}
&\epsilon\approx1\mathrm{-}\prod_{j\mathrm{=}2}^{N}{\left( 1\mathrm{-}\frac{j\mathrm{-}1}{b\mathrm{\times}2^f} \right)}\\
\end{aligned}\end{equation}

Following the similar estimation as above, we formalize the relationship between fingerprint field length $f$ and the target false positive rate $\epsilon$:
\begin{equation}\begin{aligned}
f\approx \log _2w\mathrm{+}\log _2\frac{1}{\epsilon}\mathrm{+}\textbf{1}
\label{diff}
\end{aligned}\end{equation}
Note that the appended one bit in each fingerprint field is the differentiated bit.

Equ. \ref{f} and Equ. \ref{diff} demonstrate that, with the same target false positive rate $\epsilon$, the length of fingerprint field $f$ will stay the same whether we introduce the differentiated bit or not. In fact, this bit acts as a part of fingerprint is more effective than acts as the differentiated bit. This phenomenon can be explained as follows. On the one hand, when the bit acts as a part of fingerprint, it affects all elements inserted into CCF. It is efficient that all elements have one more bit to distinguish with each other. On the other hand, when the bit is used as the differentiated bit, the bit is functional only when the conflicted tuples like $(1,2,\eta)$ and $(2,1,\eta)$ occur. However, the probability of such order of bucket indexes is negligible. Therefore, we leverage the additional bit as a part of element fingerprint.

Conclusively, to get the false positive rate decreased by $38.15\%$, CBF needs to add one more counter for each distinct element in the multiset. By contrast, CCF can get the false positive rate decreased by $50\%$ by only augmenting one more bit to each fingerprint. Compare with CBF, to obtain the targeted $\epsilon$, CCF is more space-efficient and communication-friendly for multiset synchronization. Furthermore, as reported in \cite{gremillion1982designing} and our later implementation, the value calculated by the Equ. \ref{cbf_false} is lower than the practice of CBF. In contrast, the value estimated by the Equ. \ref{false_positive} is a precise estimation of the false positive rate for CCF.

\subsection{Computation overhead}

The most computation-intensive operation is the hash calculation during the synchronization process. For each element, CBF needs to generate $k$ hash values to map the element into the vector. To ensure a decreasing false positive rate, CBF needs to lengthen the vector and implement more independent hash functions. However, CCF only requires two hash functions: one for fingerprint generation and another for the candidate bucket calculation. Its false positive rate is only decided by the bucket amount $b$ and fingerprint length $f$, thus has nothing to do with the number of hash functions.

Furthermore, in multiset synchronization, deletion and query are employed to identify the data content and get the multiplicity of corresponding element. CBF needs to get $\lceil \frac{m}{n}\ln 2 \rceil$ hashes to locate the counters for each element and get the minimum number of counters. However, in CCF, there are only two fixed mapping processes during deletion and query.

\subsection{Query-based v.s decoding-based method of CCF}

As stated in Section 3, CCF can identify the different elements via either a query-based method or a decoding-based method. In this subsection, we compare them theoretically.

The query-based method queries all the local elements against the received CCF from the other host to decide the different elements. By doing so, CCF can identify the elements in $d_M$ and generate the given number of replicas to complete the synchronization mission. The elements in $d_E$, by contrast, will be sent to the other host for synchronization. Therefore, the time-complexity of this method is $O(N_A\mathrm{+}N_B)$, where $N_A$ and $N_B$ are the number of elements in root set $A^*$ and $B^*$, respectively.

The decoding-based method tries to fetch the different element fingerprints from the decoded CCF. But it is still tricky to recover the real element content from the decoded element fingerprints. If the length of fingerprint is large enough to identify the element clearly, we can maintain an ID table to record the mapping between the fingerprint and the corresponding element content. Such table helps to retrieve the element content reversely and easily.  However, if we use fingerprint as ID to distinguish the elements from each other, the space overhead of the fingerprint field for each element is $\log \left(N\right)$ bits, which is beyond our desired. Moreover, if we decrease the number of bits for the fingerprint field, there will be much more collisions. Especially, if we use ID table to identify the elements' content, there will be $\frac{N}{2^f}$ $(\frac{b\times w}{2^f})$ elements corresponding to one fingerprint at most (suppose the $2^f$ fingerprints are mapped uniformly).  Even if $\frac{N}{2^f}\mathrm=1$, one fingerprint is corresponding to only one element, there will be much more communication overhead compared with query-based method. To avoid this overhead, we take the decoding strategy to eliminate the common elements and the query strategy to identify the different elements.

In conclusion, CCF outperforms CBF in terms of synchronization accuracy and space-efficiency. For multiset synchronization, the time-consumption of CCF is only decided by the number of distinct elements in the multiset. But for CBF, the time-consumption is jointly determined by the value of $N$ and the target false positive rate $\epsilon$.

\section{Evaluation}

In this section, we implement both the CBF-based and the CCF-based methods to compare their synchronization accuracy and time-consumption. The synchronization accuracy is defined as the percentage of common elements after synchronization:
\begin{equation}\begin{aligned}
\alpha =\frac{|A\cap B| }{|A\cup B| }
\end{aligned}\end{equation}
We also quantify the respective time-consumption of element insertion, query and deletion in both CCF and CBF.

\begin{figure}
\centering
\subfigure[Accuracy when bpe varies] {\includegraphics[width=2.3in]{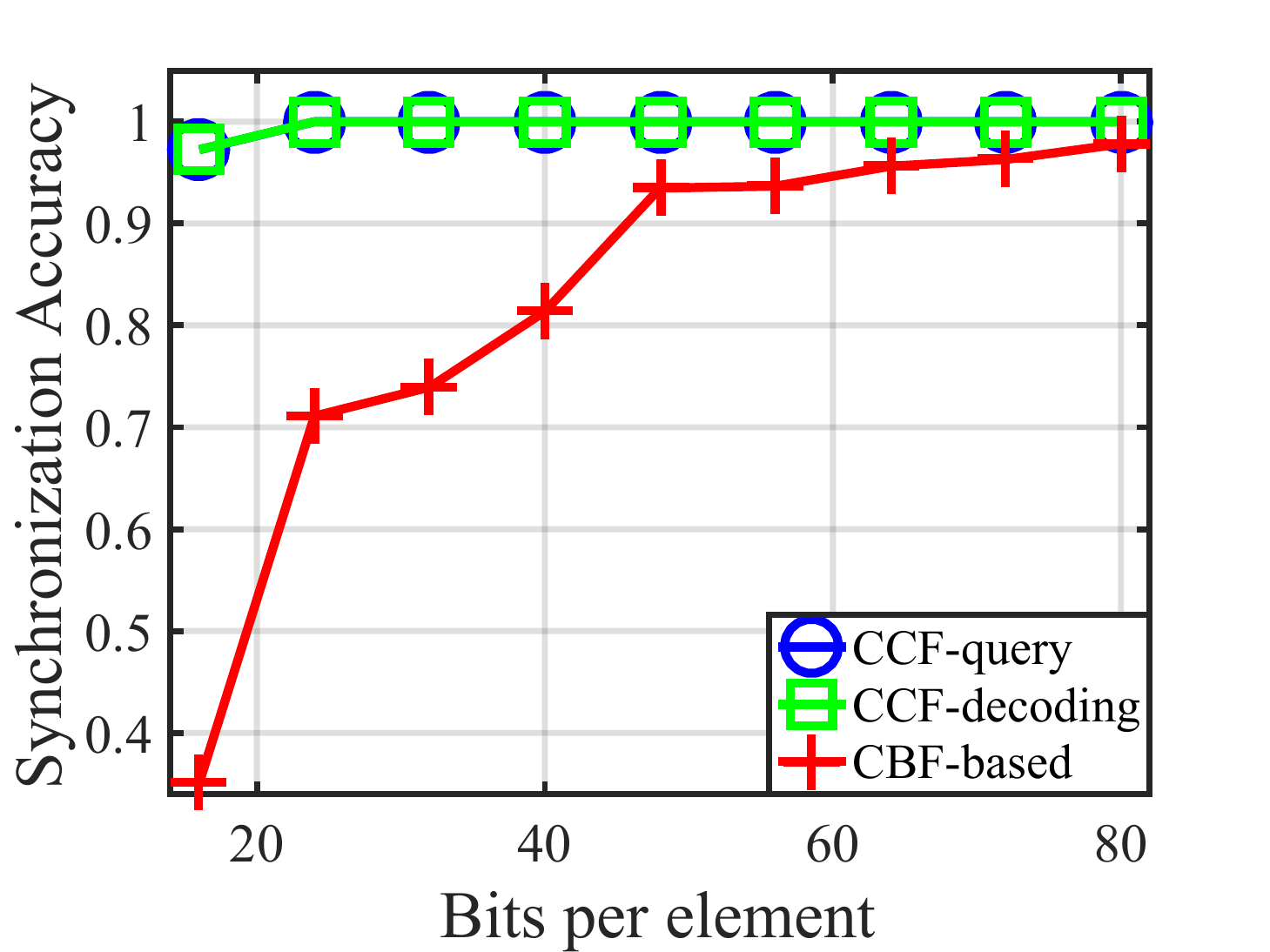}}
\subfigure[Bpe when accuracy varies] {\includegraphics[width=2.3in]{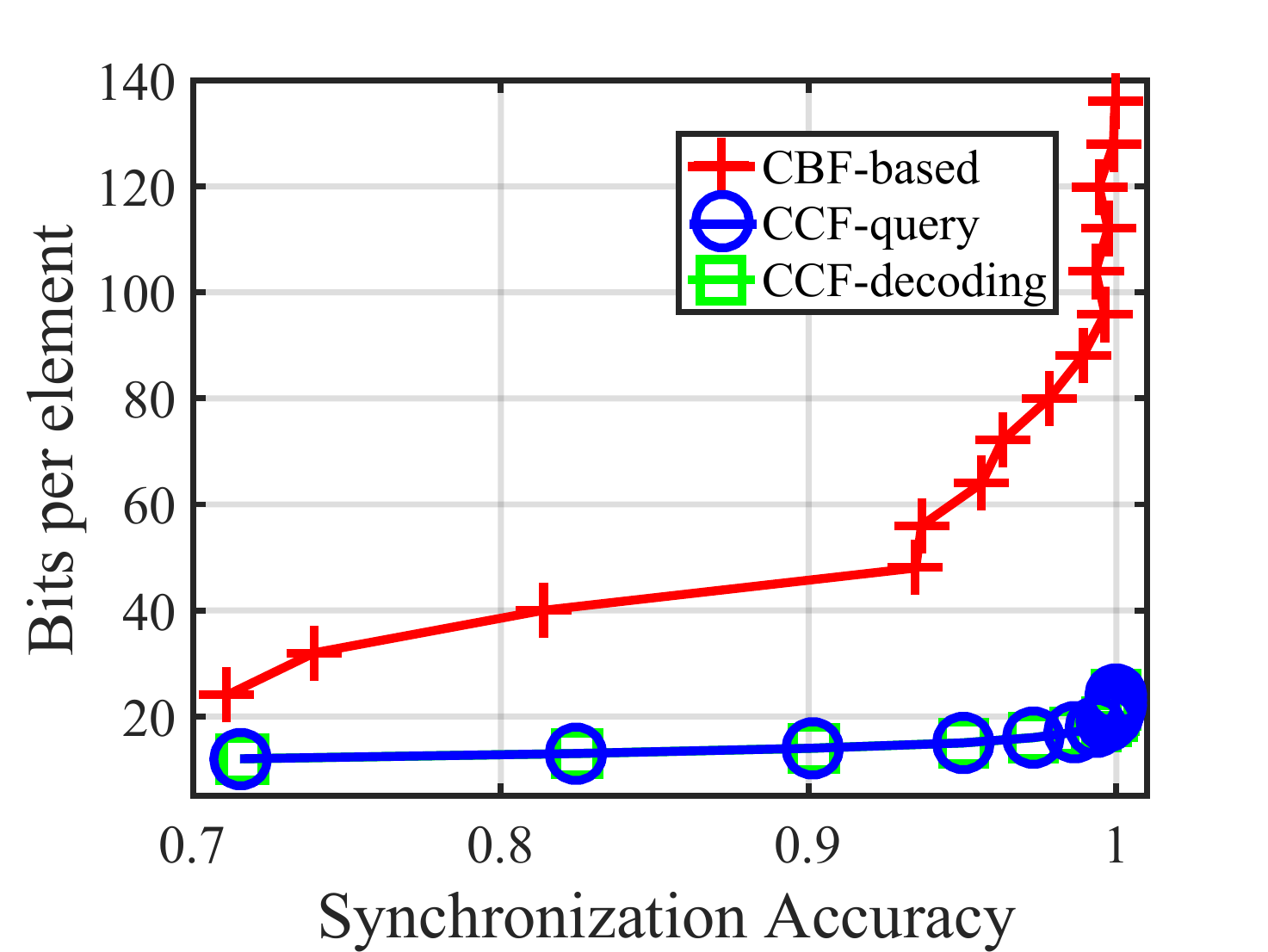}}
\subfigure[Insertion time consumption] {\includegraphics[width=2.3in]{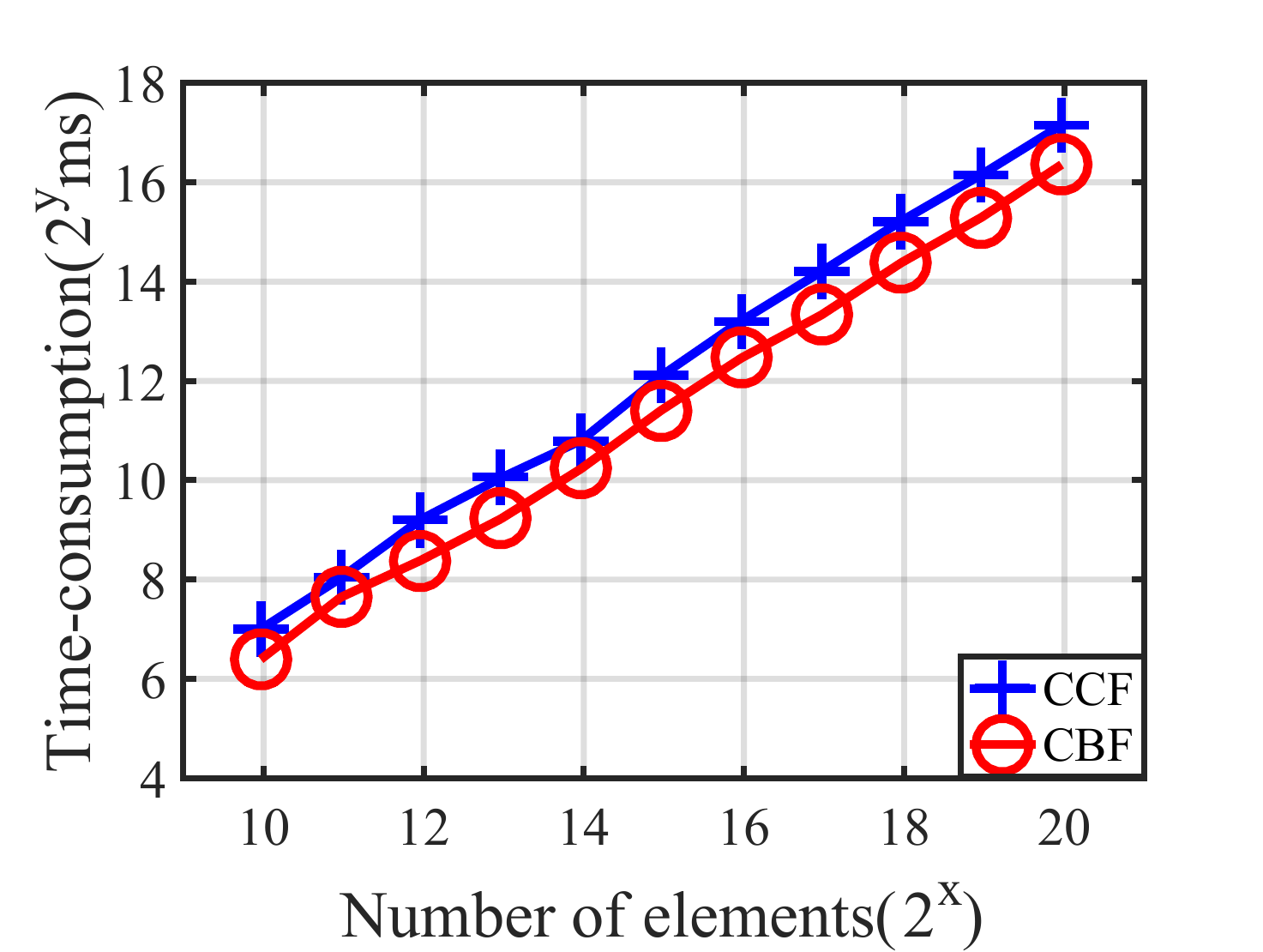}}
\subfigure[Identify different elements] {\includegraphics[width=2.3in]{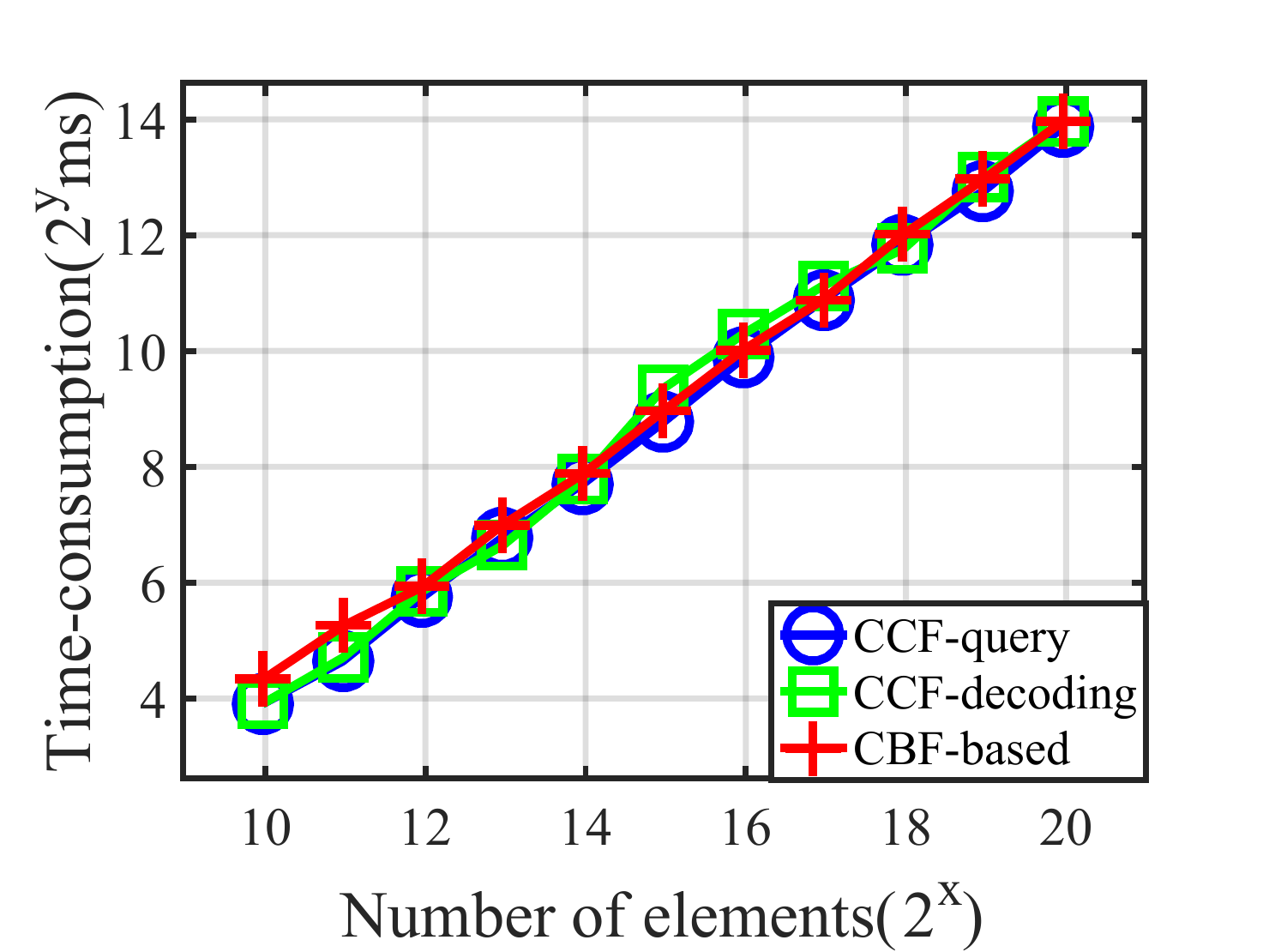}}
\caption{Performance comparison between CCF and CBF}
\label{CCCB}
\end{figure}

\subsection{Experiment methodology}
A testbed with 2.24GHz CPU and 8GB RAM is employed as a host. All the multiset elements are 32-bit integers derived out by a random number generator. For the CBF-based method \cite{guo2013set, luo2017efficient}, we borrow the CBF implementation from one of the prior work \cite{DBLP:conf/esa/KirschM06}. The independent $k$ hash functions for a CBF with $m$ cells are generated as follows:
\begin{equation}\begin{aligned}
h_i\left( x \right) \mathrm{=}\left( g_1\left( x \right)\mathrm{+}i\mathrm{\times}g_2\left( x \right) \right) \ mod\ m
\end{aligned}\end{equation}
where $g_1(x)$ and $g_2(x)$ are two random and independent integers ranging from 1 to $m$. The integer $i$ belongs to the range $[0,k\mathrm{-}1]$. As for CCF, we need two independent hash functions, one for the fingerprint generation and another for the candidate bucket selection. We let the threshold of reallocation times $max$ equal to bucket number $b$ instead of fixed $500$ in \cite{fan2014cuckoo}. The number of slots in each bucket $w$ is fixed as 4. As for hash functions, we choose CityHash\cite{timmurphy.org} for each element to generate the random candidate buckets index and the fingerprint.

\subsection{Comparing CCF with CBF}

In this subsection, we compare the performance of CCF and CBF in terms of space-efficiency, time-consumption and synchronization accuracy.

We first consider the relationship between space overhead and the target false positive rate in CCF and CBF. In our setting, the multiplicity of each element is no more than 255, so each counter in CCF with 8 bits is enough to ensure that the counters will not overflow. However, for CBF, 8-bit counters may still incur the overflow problem. The reason is that, CBF adds the corresponding $k$ counters up whenever an element is inserted. Therefore, in case of overflow, CBF has to oversubscribe the number of bits in each counter. As for the false positive rate of membership query, in CBF, $\epsilon\mathrm{\approx}0.6185^{\frac{m}{n}}$; while for CCF, $\epsilon\mathrm{\approx} \frac{w}{2^{f-1}}$. Intuitively, CBF has to occupy more space to achieve similar false positive rate decrements than our CCF data structure.

We then vary the bits per element (bpe) in CCF from 16 to 80. For CCF, more bits are used to augment fingerprint $\eta$; for CBF, the added bits are used to lengthen the CBF vector to represent the elements content and multiplicity information. And we maintain the number of elements in each root set as fixed 64000. The cardinality of each multiset is fixed as 640000, which is 10 times as many as the number of elements in root set. As illustrated in Fig. \ref{CCCB}(a), with the changing of bits per element, the performance of CCF-based methods are more robust than the CBF-based method. For CBF, when more counters are added into the vector, the stored multiplicity information is more precise. For CCF, the more bits are used in fingerprints, the less false positive errors will occur. As demonstrated in Fig. \ref{CCCB}(b), with much less bits per elements, CCF-based methods can achieve the same synchronization accuracy as the CBF-based method. And the increased bits per element for the CCF-based methods is only 7 bits, compared with 96 bits for the CBF-based method, when the synchronization accuracy increases from 0.7 to 0.9999. Therefore, compared to the CBF-based method, given then same bpe, our CCF-based method achieves much higher synchronization accuracy; to realize the same synchronization accuracy, our CCF-based method needs much less bpe.

In our multiset synchronization framework, the hosts need to insert their local elements into the CCFs. The query-based method has to query each local element to distinguish the different elements from the common ones. The decoding-based method needs to delete common elements from the CCF vector. Therefore, we further compare the time-consumption that caused by element insertion, and different element uncovering. Certainly, the transmission of the employed data structures and the different elements in $d_E$ also consumes some time. However, this kind of time-consumption is beyond the scope of our consideration. In our experiments, we vary the number of elements in the multiset from $500\times2^1$ to $500\times2^{11}$ and record the two kinds of time-consumption in Fig. 4(c), and Fig. 4(d) respectively.

As illustrated in Fig. \ref{CCCB}(c) and Fig. \ref{CCCB}(d), CCF causes comparable (actually a little bit more) time-consumption as CBF. Note that, both CCF and CBF need to calculate two hash values when inserting an element. Specifically, CBF has to calculate two random value $g_1(x)$ and $g_2(x)$ to generate the $k$ hash values. The CCF calculates the index of the two candidate buckets for each element during element insertion, query and deletion. Besides, CCF maps the original element content to an integer in $[0, 2^f-1]$ to generate the fingerprint of each element. This explains why CCF results in a bit more time-consumption than CBF. Moreover, for element insertion, the reallocation process may be triggered in CCF, when the two candidate buckets are both occupied. This reallocation process surely increases the time-consumption of CCF insertion.

Finally, we quantify the time-consumption of identifying the different elements with CCF and CBF. During a membership query, CBF needs to checks the $k$ corresponding counters; while CCF may access the number of slots from $1$ to $2w$ randomly. Usually, $2w$ is larger than the value of $k$. The results are given in Fig. \ref{CCCB}(d). We consider both the query-based and the decoding-based methods enabled by CCF. Notice that, CBF performs better than the CCF-decoding to some extent, but slightly worse than the CCF-query. The reason is that the decoding-based method consumes more time to query the corresponding different elements after the elimination phase.

According to the above results, we conclude that, CCF achieves much better synchronization accuracy than CBF, with a little compromise of time-consumption.

\subsection{Impact of parameters in CCF}

Due to the unavoidable hash collisions, CCF incurs false positive errors of membership query. As we have analyzed in Section 5, the false positive rate is jointly determined by the bucket number $b$, fingerprint length $f$, and slot number $w$ in each bucket. The hash collisions are proportional to the synchronization accuracy of CCF. Thus in this subsection, we first analyze the false positive errors when inserting elements into CCF, then compare the generated synchronization accuracy when the query-based and decoding-based methods are employed, respectively.

\begin{table}[]
\caption{Space occupancy ratio when the number of elements varies}
\centering
\begin{tabular}{cc|cc}
\hline
\multicolumn{1}{l}{$|X|$ ($10^3$)} & \multicolumn{1}{l|}{Occupancy} & \multicolumn{1}{l}{$|X|$ ($10^3$)} & \multicolumn{1}{l}{Occupancy} \\ \hline
1 & 0.96387 & 2 & 0.96631 \\
4 & 0.96875 & 8 & 0.96683 \\
16 & 0.97004 & 32 & 0.96992 \\
64 & 0.96905 & 128 & 0.96946
\\
256 & 0.96913 & 512 & 0.96938 \\
1024 & 0.96916 & 2048 & 0.96926 \\ \hline
\label{occup}
\end{tabular}
\vspace{-4ex}
\end{table}
\begin{figure*}
\centering
\subfigure[Collisions when $b$ varies] {\includegraphics[width=2.3in]{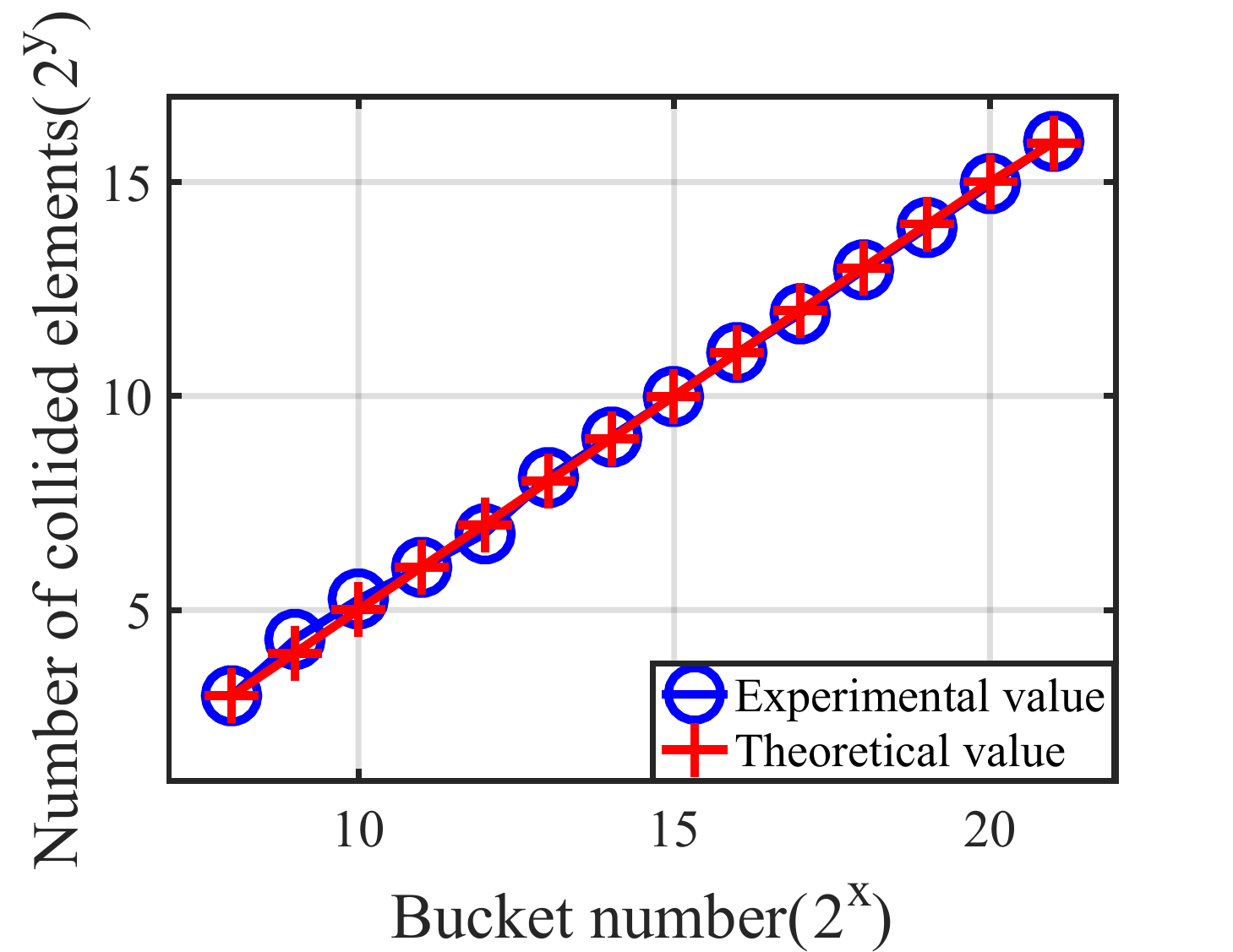}}
\subfigure[Collisions when $f$ varies] {\includegraphics[width=2.3in]{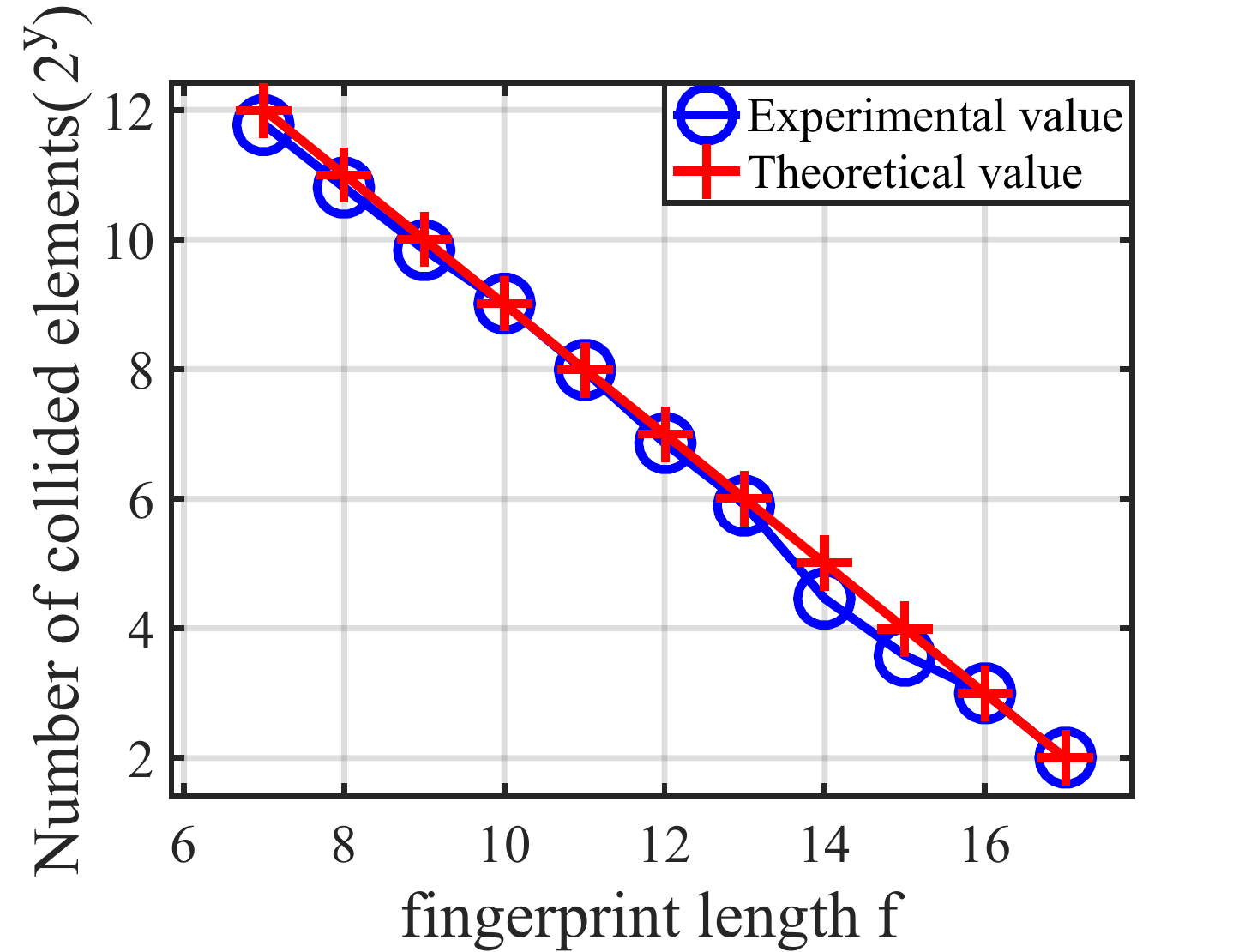}}
\subfigure[Synchronization accuracy when $|X|$ varies] {\includegraphics[width=2.3in]{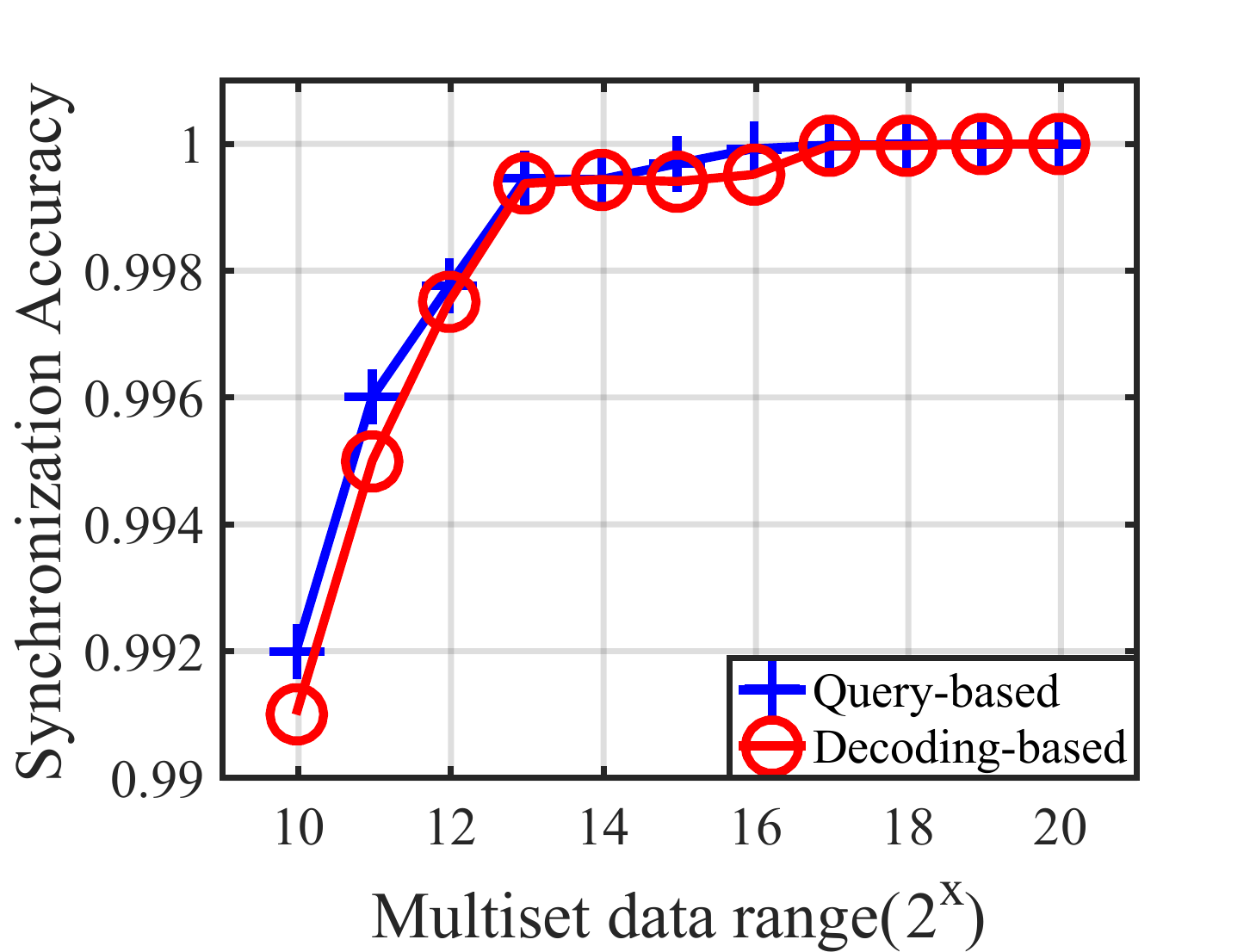}}
\subfigure[Synchronization accuracy when $f$ varies] {\includegraphics[width=2.3in]{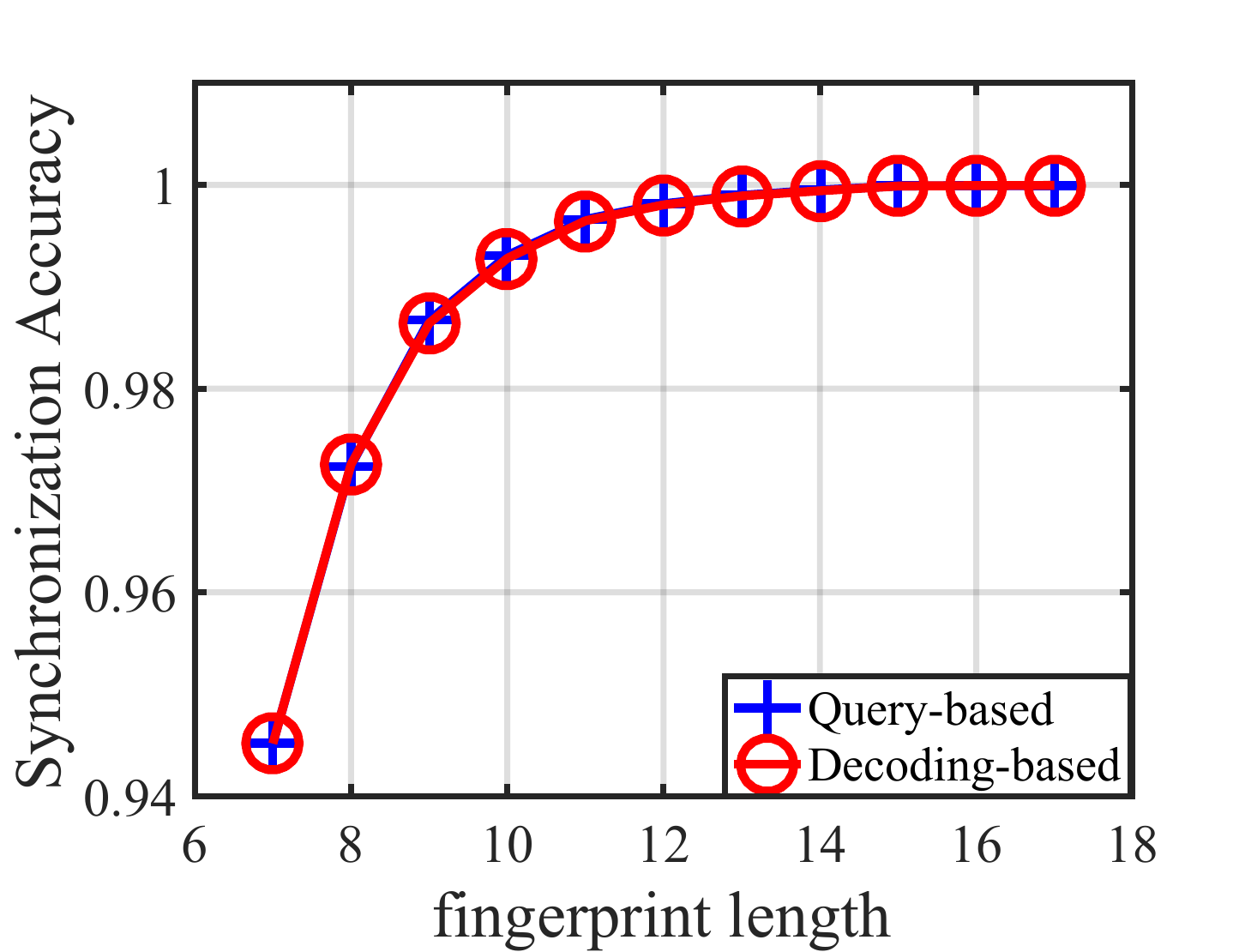}}
\caption{The performance of CCF with diverse parameter settings}
\label{accurac}
\end{figure*}
As demonstrated in Table \ref{occup}, we increase the number of buckets from $2^8$ to $2^{21}$ and change the number of distinct elements in multiset range from $2^1\mathrm{\times}500$ to $2^{14}\mathrm{\times}500$. The parameters of the CCF are set as $w\mathrm{=}4$, $f\mathrm{=}log_2b$. The experimental results show that all the multisets can be successfully represented by CCF. The ideal occupancy ratio is $500\mathrm{\div}2^7\mathrm{\approx}0.9765$. Apart from the factor of collision and random integers generation, our CCF can achieve quite high space occupancy ratio from 0.965 to 0.97. Thus in this paper, we don't focus on space occupancy ratio, only consider space-efficiency, i.e., how effectively the bits are used to control false positive rate.

We vary the parameters of CCF to quantify their impact to the CCF performance. The results are depicted in Fig. \ref{accurac}. Given $f\mathrm{=}10$, $w\mathrm{=}4$, we vary the length of CCF, i.e., $b$, from $2^8$ to $2^{21}$ and record the number of collided elements. The theoretical value of collided elements ratio is given as Equ. \ref{false_positive}. As shown in Fig. \ref{accurac}(a), the number of false positive errors is proportional to the value of $b$. With more buckets introduced, the fixed fingerprint is not enough for the elements to distinguish from each other. The number of false positive errors increases with the bucket number linearly. In contrast, as shown in Fig. \ref{accurac}(b) and Equ. \ref{f}, the increase of $f$ will significantly reduce the false positive rate.

Then we measure the impact of the number of distinct elements in the multisets and the fingerprint length $f$ to the CCF synchronization accuracy. We also consider both the query-based and decoding-based synchronization methods. As illustrated in Fig. \ref{accurac}(c), we vary the number of distinct elements from ${2^1\mathrm{\times}500}$ to ${2^{11}\mathrm{\times}500}$, the corresponding number of buckets ranges from ${2^8}$ to ${2^{18}}$ to ensure high occupancy ratio, and the corresponding fingerprint length ranges from 8 to 18 bits. As shown in Fig. \ref{accurac}(c), with such a parameter setting, the synchronization accuracy increases significantly. The reason is that, as Equ. \ref{false_positive} demonstrated, with the same number of bits used for bucket index and the fingerprint field, the number of collided elements is fixed as $b\mathrm{\times} w\mathrm{\times} \frac{w}{2^{f\mathrm{-}1}}\mathrm{=}2w^2\left( b\mathrm{=}2^f \right) $. The false positive rate will decrease with the increasing number of elements to be represented. The decoding-based method achieves a bit lower synchronization accuracy than the query-based method. For the query-based method, collisions occur only in the query phase. However, for the decoding-based method, there can be false positive errors during both the decoding and identifying phases. More false positive errors are introduced into the decoding-based method. Lastly, Fig. \ref{accurac}(d) depicts the synchronization accuracy when the fingerprint length $f$ increases from 7 to 17. Clearly, both the query-based and decoding-based methods lead to increasing synchronization accuracy (from 0.945 to 0.99999).

We conclude here that the parameters have diverse impact on the performance of CCF, and the false positive rate that we estimate can correctly evaluate the number of collided elements. The query-based method outperforms the decoding-based method in terms of synchronization accuracy.
\section{Conclusion}
In this paper, we give a detailed formulation of CCF, a novel variant of cuckoo filter, to represent and thereafter synchronize multisets. CCF extends each slot as two fields, the fingerprint field and the counter field. The fingerprint field records the element fingerprint that stored by this slot; while the counter field counts the multiplicity of the stored element. With such a design, CCF is competent to represent multisets. After exchanging the respective CCFs which represent the local multisets, the hosts determine the different elements with either the query-based or the decoding-based method. CCF is able to distinguish the different elements in $d_M$ from $d_E$, so that the elements in $d_M$ can be synchronized by generating the dedicated number of local replicas. Only the different elements in $d_E$ need to be transmitted to the other host with multiplicity information. This property decreases the communication overhead significantly. The comprehensive evaluation results indicate that CCF outperforms CBF in terms of synchronization accuracy and space-efficiency, at the cost of a little higher time-consumption.

\bibliographystyle{plain}
\bibliography{CCF}

\end{document}